\documentclass[preprint,aps,amsmath,nofootinbib,tightenlines]{revtex4}
\newcommand{\nn}{\nonumber}

\usepackage{epsfig}
\usepackage{subfigure}

\begin{document}
\preprint{\vbox{\hbox{CALT-68-2778}}}

\title{\phantom{x}
\vspace{0.5cm}
Solitonic Brane Inflation \\[6pt] 
\vspace{0.5cm}
}

\author{Kevin T. Engel}
\affiliation{California Institute of Technology,
Pasadena, CA 91125\footnote{Electronic address: kte@caltech.edu}
\vspace{0.2cm}}

\begin{abstract}
\vspace{0.3cm}

We present a new type of brane inflation motivated by multi-kink solitonic solutions of a scalar field in five dimensions.  In the thin brane limit, we analyze a non-static configuration in which the distance between two parallel domain walls decreases.  We show that the ensuing spacetime is inflationary, both on the branes, and, for certain potentials, in the bulk.  We argue that this inflationary regime is transitory and can end via a brane merger into a single kink solution - a flat, thick brane RS2 universe.  This scenario is quite general; we show that any potential which supports a single flat kink solution is also likely to support an inflationary multi-kink configuration.

\end{abstract}

\maketitle

\section{Introduction}
Extra-dimensional theories in which our universe contains more than the usual three spatial dimensions have enjoyed a resurgence in the last few decades .  Traditionally the extra dimensions are assumed to be small so that their effects can be hidden through a Kaluza-Klein compactification.  Recently, though, new ideas have emerged in which standard model particles are confined to a hypersurface (brane) which is embedded in a larger dimensional bulk region \cite{Rubakov:1983bb,Horava:1995qa,ArkaniHamed:1998rs,Randall:1999ee}.  One interesting realization of these ideas is the model proposed by Randall and Sundrum which has come to be known as the RS2 universe \cite{Randall:1999vf}.  In this model, our observable universe is contained on a brane with positive tension $\Lambda_4$ located at $z=0$ in the additional dimension.  If the bulk region is dominated by a negative cosmological constant which satisfies the fine tuned relationship: $\Lambda_5 = -\Lambda_4^2/(6M^3)$, where $M$ is proportional to the 5D Planck mass, then a static but warped spacetime exists with line element:
\begin{equation}
ds^2 = e^{-\Lambda_4 \left|z\right|/(6M^3)} \eta_{\mu\nu}dx^\mu dx^\nu+dz^2\,.
\end{equation}
In flat D-dimensional spacetime, the gravitational force falls off as $r^{2-D}$, but Randall and Sundrum showed that near the brane, the warped space effectively localizes gravity in the extra dimension, thus reproducing the familiar $1/r^2$ gravitational force.

In this braneworld, like most others, the effects of the extra dimensions become more pronounced at higher energies.  As our universe was once much hotter, the cosmology of these extra dimensional scenarios requires some study (For a review, see Refs. \cite{Brax:2003fv,Langlois:2002bb}).  In particular, extra dimensional theories have given rise to new methods for generating inflation.  In the simplest models the inflaton is a field confined to the brane \cite{Kaloper:1998sw,Maartens:1999hf}; these models resemble the standard slow roll inflation but with less restrictive slow roll conditions.  Others have considered a bulk inflaton \cite{Mohapatra:2000cm,Maeda:2000wr, Himemoto:2000nd}.  A more drastic departure from the standard lore was proposed by Dvali and Tye in which inflation is induced from the motion of the brane through the bulk \cite{Dvali:1998pa}.  Two of these objects eventually collide, ending inflation and reheating the branes.  Brane collisions also take place in ekpyrotic cosmologies, which has been presented as an alternative to inflation \cite{Khoury:2001wf}.  Models such as these are often motivated by string theory.

In this paper, we consider a new model of brane inflation motivated instead by solitonic solutions in quantum field theory.  Initially idealized as infinitely thin delta function terms in the action, there has been considerable interest recently in explaining branes in terms of solitonic solutions of a bulk scalar field (For a review, see Ref. \cite{Dzhunushaliev:2009va}).  These thick brane solutions offer a simple source for the brane term which is inserted by hand in the RS2 model.  Thick branes also demonstrate the required ability to trap particles, as matter fields in the presence of the solitonic background have been shown to possess eigenstates localized around the brane \cite{Rubakov:1983bb,
Dvali:1996xe,Bajc:1999mh,Slatyer:2006un}  

\begin{figure}[] 
\subfigure[~scalar potential]{
\includegraphics[scale = .48]{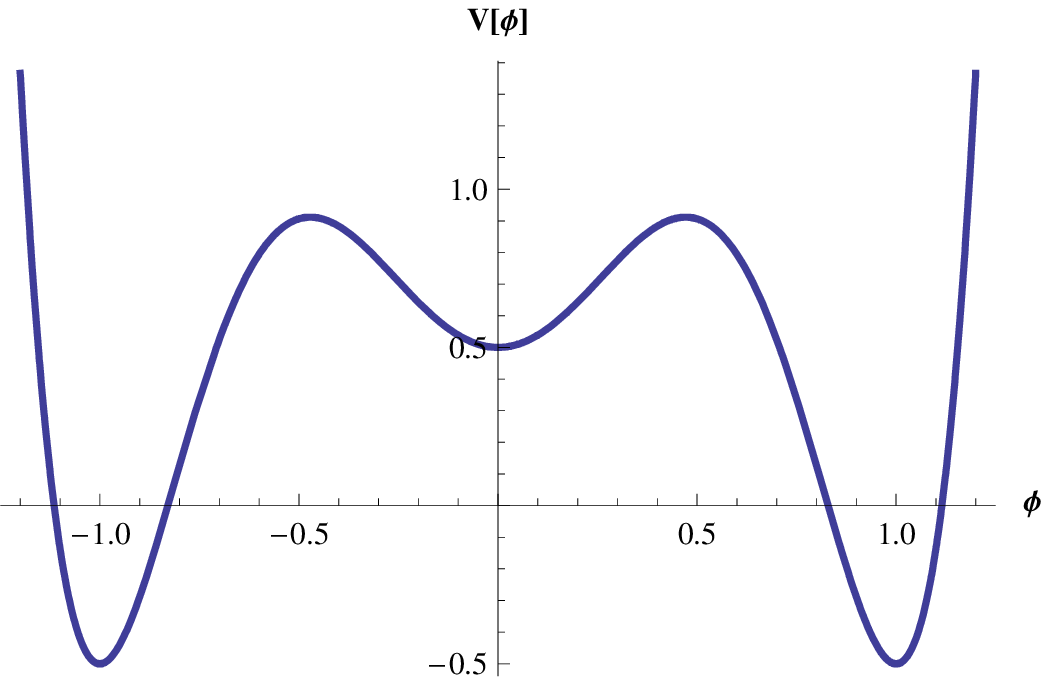}
\label{pot1}
}
\subfigure[~kink-kink initial state]{
\includegraphics[scale = .48]{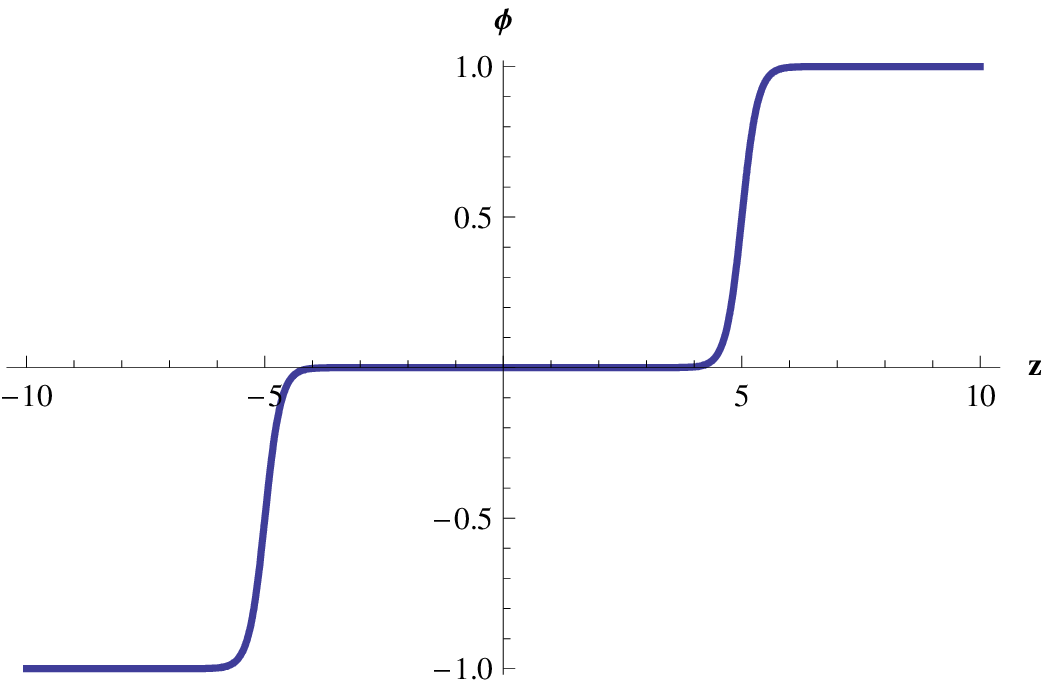}
\label{kk}
} 
\subfigure[~kink final state]{
\includegraphics[scale = .48]{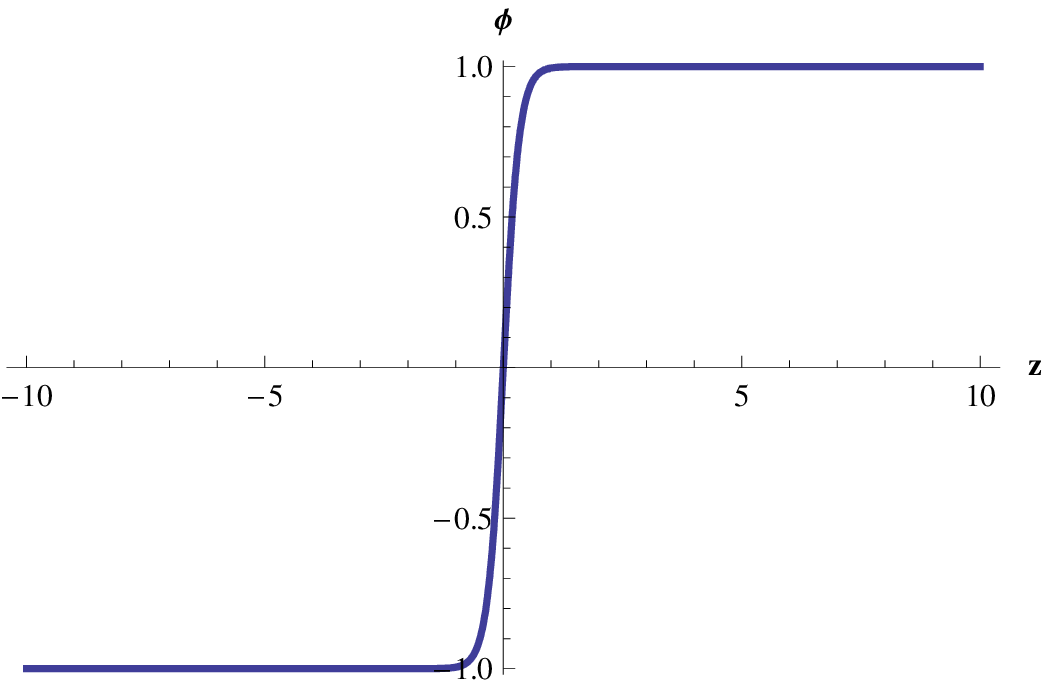}
\label{k1}
}
\subfigure[~scalar potential]{
\includegraphics[scale = .48]{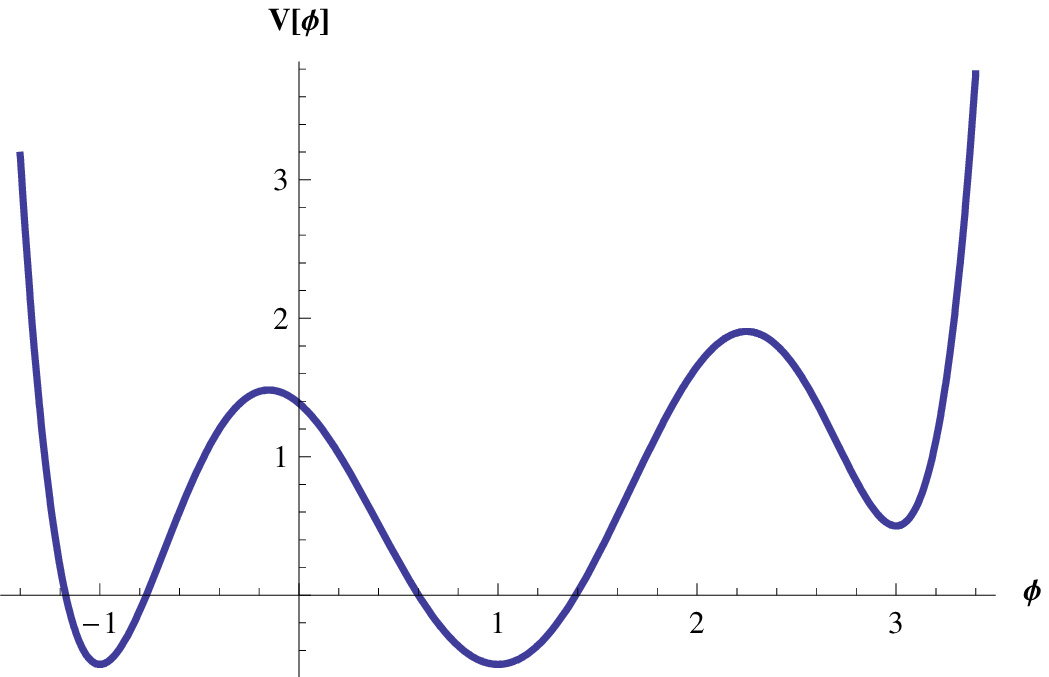}
\label{pot2}
}
\subfigure[~kink-antikink initial state]{
\includegraphics[scale = .48]{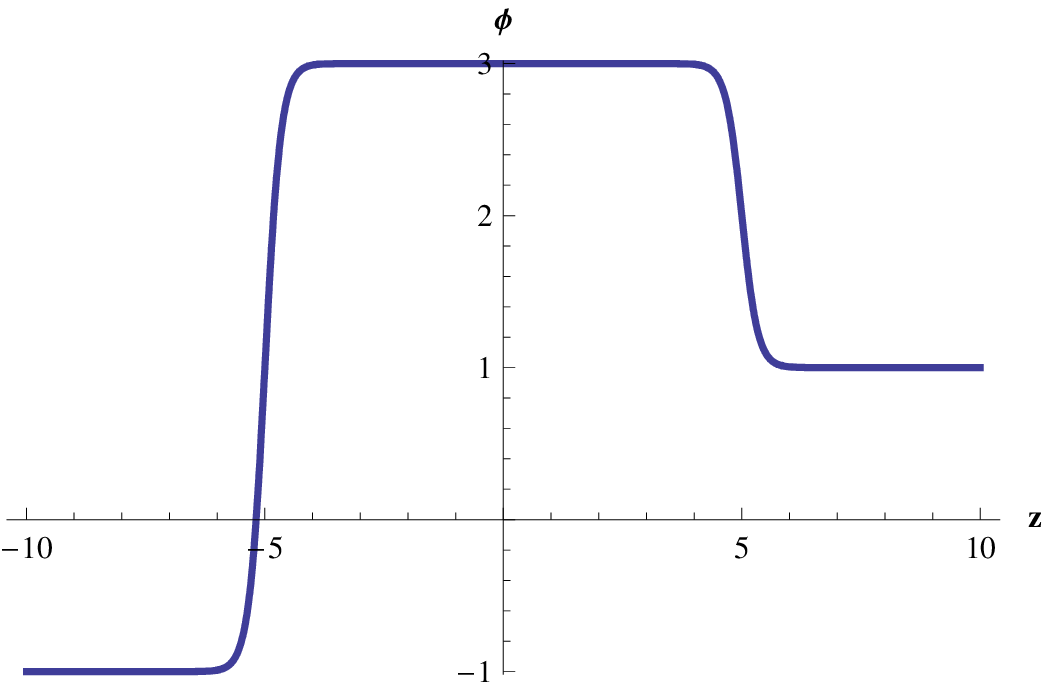}
\label{kak}
}
\subfigure[~kink final state]{
\includegraphics[scale = .48]{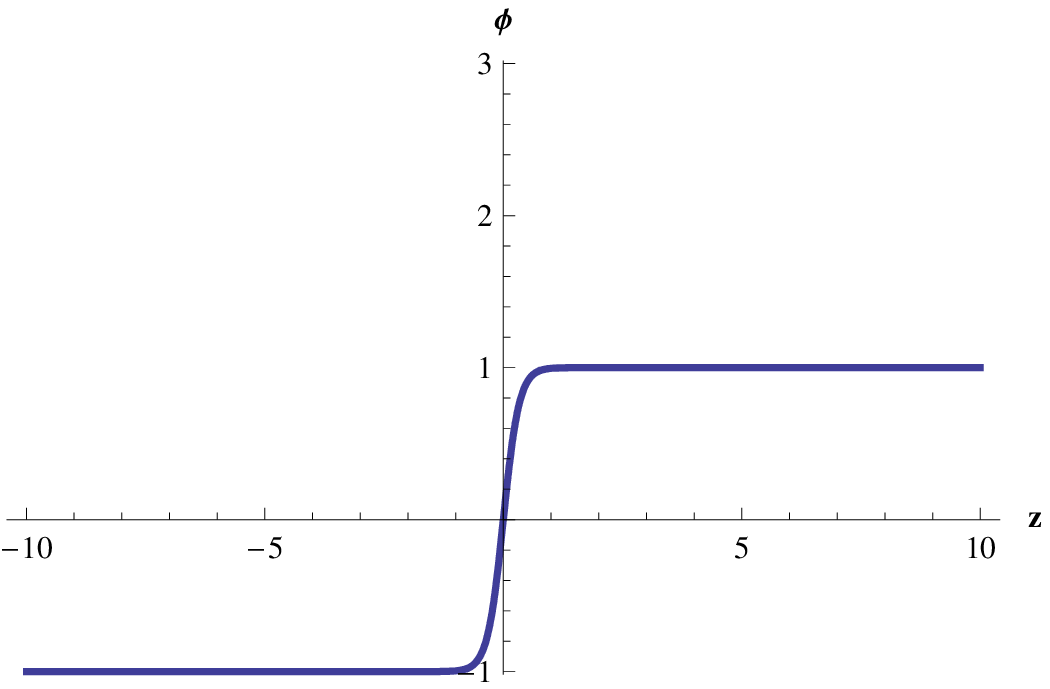}
\label{k2}
}
\caption{Two potentials capable of supporting multi-kink states which may evolve into a single kink final state.}
\label{profiles}
\end{figure}

In this work, we present another advantage, arguing that a natural inflationary regime exists within the thick brane framework - a solitonic analogue of Dvali and Tye's brane inflation.  In Minkowski space, a potential which supports a single kink solution also supports multi-kink solutions.  Similarly, we expect that a potential capable of producing a thick brane will also support multi-brane configurations.  Just as in the RS2 case, a tuned potential is required to produce a flat brane, therefore a multi-brane configuration will generically consist of branes whose internal geometry is either de Sitter or anti-de Sitter.  For field configurations which end in the appropriate vacua, we will show that the de Sitter multi-brane states provide a simple inflationary extension of the single thick brane universe.  We consider here only the simplest possibility in which two inflationary branes merge to form the single flat RS2 brane.  Schematic configurations for the scalar field and its potential are shown in Figure \ref{profiles}.  Note that both initial configurations require a potential with at least 3 local minima.  This scenario still works with a double well potential, but in order for the final state to exist, the initial state must contain an odd number of branes.

The brane inflation presented here is original in its application to multi-kink solitonic solutions, however, one other inflationary scenario deserves mention as it, too, produces a thick brane RS2 universe.  In Ref. \cite{Bucher:2002np}, Bucher proposed an initial false vacuum de Sitter phase which decays through bubble nucleation into a true vacuum AdS phase.  If degenerate AdS vacua exist, then a bubble collision will result in the formation of a brane at the interface.  However, susequent calculations of the cosmological perturbations in this scenario disagree with observations as they predict a red spectral index $n_s > 1$ \cite{BlancoPillado:2001si,Garriga:2001qn}.  Although we do not calculate the cosmological perturbations for our solitonic model, enough differences exist at the classical level to suggest that a different result may be obtained.

For the remainder of the paper we will analyze the period of solitonic inflation in the thin brane limit in which the width of the branes goes to zero.  In Section \ref{thinbranes} we find the spacetime geometry induced by two thin branes moving together.  Our results appear to be original as most of the thin brane solutions in the literature are static ones.  Some non-static solutions are presented in Ref. \cite{Tziolas:2007di}, but their branes are not the standard thin ones which are described solely by their tension.  In Section \ref{inflation} we discuss the inflationary aspects of the two brane spacetime which solve the same problems as standard 4D inflation.  Next, in Section \ref{branemerger}, we argue that under certain circumstances, the two branes will merge to form the desired final state of a flat, single brane RS2 universe.  In Section \ref{finetuning} we address issues of fine tuning, showing that the only required tuning in this scenario is the standard one to set the cosmological constant on the final RS2 brane equal to zero.  Finally in Section \ref{conclusion} we summarize our results and point out further lines of inquiry suggested by this work.

\section{Thin branes}\label{thinbranes}
Consider a pre-RS2 universe containing two thin branes.  Anticipating an eventual merger, we require they be parallel and the distance between them decreasing.  If coordinates are chosen so that the branes move together at a constant velocity v, then the thin brane action is given by:
\begin{equation} \label{action1}
S = -\int \sqrt{-g}\, \Lambda_5(t,z)-\int \sqrt{-g^{\rm ind}}\,\frac{\Lambda_{4-}}{\gamma}\delta(z-vt)-\int \sqrt{-g^{\rm ind}}\,\frac{\Lambda_{4+}}{\gamma}\delta(z+vt) \,,
\end{equation}
where $\gamma$ (inserted for later convenience) is the usual $1/\sqrt{1-v^2}$ and $\Lambda_{4\pm}$ are the brane tensions.  Each brane has an associated 4D metric 
$g^{\rm ind}$ whose embedding in five dimensions is given by:
\begin{equation}\label{indmetric}
g^{\rm ind}_{AB} = g_{AB}-n_An_B \,,
\end{equation}
where $n_A$ is the unit normal to the brane.  In this paper we use uppercase Latin indices to represent the full 5D coordinates (0,1,2,3,5), greek ones to represent 4D brane coordinates (0,1,2,3), and lowercase Latin ones for the three spatial coordinates (1,2,3) on the brane.  In a thick brane scenario, $\Lambda_5$
corresponds to the value of the scalar potential at a local minima; since kinks interpolate between different minima, we allow $\Lambda_5$ to take arbitrary values between branes:
\begin{equation}
\Lambda_5(t,z) = \Lambda_5^{\rm I}\theta(-z+vt)+\Lambda_5^{\rm II}\left(
\theta(z-vt)-\theta(z+vt)\right)+\Lambda_5^{\rm III}\theta(z+vt)\,.
\end{equation}
From this action the stress energy tensor can be calculated:
\begin{equation}
T_{AB} = \frac{-2}{\sqrt{-g}}\frac{\delta S}{\delta g^{AB}} = 
-g_{AB}\Lambda_5(t,z)-g_{AB}^{\rm ind}\frac{\sqrt{-g^{\rm ind}}}{\sqrt{-g}}
\frac{\Lambda_{4-}}{\gamma}\delta(z-vt)-g_{AB}^{\rm ind}\frac{\sqrt{-g^{\rm ind}}}{\sqrt{-g}}\frac{\Lambda_{4+}}{\gamma}\delta(z+vt) \,.
\end{equation}
We anticipate a potential objection to this setup, namely that the motion of the branes is not being treated in a dynamic manner.  However, since any monotonically increasing trajectory can be recast through a coordinate transformation into a constant velocity solution, we are choosing to work in the coordinate system where the brane dynamics are trivial.  We assauge any doubts in the appendix by repeating the calculations presented in this section but with two brane positions which are dynamic variables.

Next we choose a parameterization for $g_{AB}$.  For the inflationary solutions we are interested in, the metric components will be independent of the three brane spatial coordinates.  With this stipulation, the most general metric is:
\begin{equation}
g_{AB} = \left(\begin{array}{ccc}
g_{00}(t,z)& &g_{05}(t,z)\\
&g_{ii}(t,z)&\\
g_{05}(t,z)& &g_{55}(t,z)\end{array}\right) \,,
\end{equation}
where the three brane spatial coordinates are represented by one entry.
If we are willing to allow for more general brane trajectories, then two 
independent coordinate transformations can simplify this parameterization:
\begin{equation}
g_{AB} = \left(\begin{array}{ccc}
-a^2(t,z)& &\\
&a^2(t,z)&\\
& &n^2(t,z)\end{array}\right) \,.
\end{equation}
This paper will focus on the restricted subset of solutions which maintain a constant velocity brane trajectory in this coordinate system and furthermore have $a = n$.  This leads to a conformally flat metric ansatz:
\begin{equation} 
g_{AB} = n^2(t,z)\eta_{AB}\,.
\end{equation}

Before solving Einstein's equations, we must first find the induced metric for each brane.  The brane on the left (for $t < 0$) has a worldline proportional to $u^A = (1,0,0,0,v)$.  Omitting the unimportant first 3 spatial coordinates, $n_A$ can be parameterized as $(n_0,n_5)$.  Since $n$ must be orthogonal to $u$ and have unit norm, we get two equations:
\begin{align}
u^An_A &= n_0+vn_5 = 0 \,, \\
g^{AB}n_An_B &= n(t,z)^{-2}(-n_0^2+n_5^2) = 1 \,, \nn
\end{align}
whose solution is given by 
\begin{equation}
n_A =  n(t,z)\gamma\left(\begin{array}{c}-v\\1\end{array}\right) \,.
\end{equation}
From Eq. \eqref{indmetric} we find the induced metric for the left brane:
\begin{equation} \label{inducedmetric}
g_{AB}^{\rm ind,L} = n^2(t,z)\left(\begin{array}{ccc}
-\gamma^2&&v\gamma^2\\
&1&\\
v\gamma^2&&-v^2\gamma^2\end{array}\right) \,.
\end{equation}
The determinant of this metric is necessary to calculate $T_{AB}$, but as an embedding of a 4D metric, one of its eigenvalues is 0.  To find the non-zero eigenvalues, we diagonalize with a Lorentz boost in the z-direction:
\begin{equation}
\left(\begin{array}{ccc}\gamma&&v\gamma\\&1&\\v\gamma&&\gamma\end{array}\right)
n^2(t,z)\left(\begin{array}{ccc}-\gamma^2&&v\gamma^2\\&1&\\
v\gamma^2&&-v^2\gamma^2\end{array}\right)
\left(\begin{array}{ccc}\gamma&&v\gamma\\&1&\\v\gamma&&\gamma\end{array}\right)
= n^2(t,z)\left(\begin{array}{ccc}-1&&\\&1&\\&&0\end{array}\right)\,.
\end{equation}
From this, we conclude that:
\begin{equation}
\sqrt{-g^{\rm ind}} = n^4(t,z)\,.
\end{equation}
The induced metric for the brane on the right is given by swapping $v\rightarrow-v$.

Finally Einstein's equations can be solved.  The non-zero stress energy components are:
\begin{align}
T_{00} &= n^2\Lambda_5+\left|n\right|\gamma\Lambda_{4-}\delta(z-vt)+ \left|n\right|\gamma\Lambda_{4+}\delta(z+vt) \,, \\
T_{ii} &= -n^2\Lambda_5-\left|n\right|\frac{\Lambda_{4-}}{\gamma}\delta(z-vt)-
\left|n\right|\frac{\Lambda_{4+}}{\gamma}\delta(z+vt) \,, \nn \\
T_{55} &= -n^2\Lambda_5+\left|n\right|v^2\gamma\Lambda_{4-}\delta(z-vt)+
\left|n\right|v^2\gamma\Lambda_{4+}\delta(z+vt) \,, \nn \\
T_{05} &= -\left|n\right|v\gamma\Lambda_{4-}\delta(z-vt)+
\left|n\right|v\gamma\Lambda_{4+}\delta(z+vt) \,, \nn
\end{align}
and the non-zero Einstein tensor components are:
\begin{align}
G_{00} &= 6\frac{n_t^2}{n^2}-3\frac{n_{zz}}{n} \,, \\
G_{ii} &= 3\frac{n_{zz}}{n}-3\frac{n_{tt}}{n} \,, \nn \\
G_{55} &= 6\frac{n_z^2}{n^2}-3\frac{n_{tt}}{n} \,, \nn \\
G_{05} &= 6\frac{n_tn_z}{n^2}-3\frac{n_{tz}}{n} \,, \nn
\end{align}
where the $t,z$ subscripts represent partial derivatives.  The most general
$n$ which satisfies these equations is given by:
\begin{equation}
n(t,z) = (b+c\left|z-vt\right|+d\left|z+vt\right|+ez+fvt)^{-1}\,,
\end{equation}
where $b$ can be chosen arbitrarily, but the other 
coefficients are constrained:
\begin{align}
c &= \frac{\gamma \Lambda_{4-}}{6M^3} \,, \label{con-1}\\
d &= \frac{\gamma \Lambda_{4+}}{6M^3} \,, \label{con0}\\
(e-c-d)^2-(-c+d-f)^2v^2 &=-\frac{\Lambda_5^{\rm I}}{6M^3} \,, \label{con1} \\ 
(e+c-d)^2-(c+d-f)^2 v^2 &= -\frac{\Lambda_5^{\rm II}}{6M^3} \,, \label{con2} \\
(e+c+d)^2-(c-d-f)^2 v^2 &= -\frac{\Lambda_5^{\rm III}}{6M^3} \,. \label{con3}
\end{align}
The $M$ in these equations is proportional to the 5D Planck mass: $M^{-3} = 8\pi G_5$.  In deriving these constraints, we have assumed that $b$ is chosen so that $n$ is positive at both branes.  If this metric solution is required to be valid for $t\rightarrow-\infty$, the presence of the $\left|n\right|$ in $T_{AB}$ enforces the additional constraint:
\begin{equation}
f < \rm{Min}(2c+e,2d-e)\,.
\end{equation}

This solution is one of the main results of this paper as it demonstrates the general existence of non-static, multi-kink configurations.  In a thick brane scenario, the various $\Lambda$'s are determined by the scalar potential.  This leaves 5 equations for the 5 unknowns $c$,$\,d$,$\,e$,$\,f$, and $v$.  Admittedly, the equations are non-linear, so a solution is not guaranteed.  However, for a set of randomly chosen $\Lambda$'s of $\mathcal{O}(M)$ which could produce a stable RS2 universe ($\Lambda_5^{\rm I} = \Lambda_5^{\rm III} < 0$,
$\Lambda_5^{\rm II}>\Lambda_5^{\rm I}$, $\Lambda_{4\pm} > 0$) a solution which satisfies all the constraints was found in roughly half the cases.  Recalling our restrictive assumptions about the brane trajectory and the form of the metric, this strongly suggests that a generic potential with three or more local minima will support both a single kink solution and a multi-kink solution. 

\section{Inflation}\label{inflation}
If our universe is a braneworld, then observations require a high degree of homogeneity and isotropy both on the brane and in the bulk at early times.  The requirement of bulk smoothness is necessary since gravity can induce brane inhomogeneities from bulk ones \cite{Bucher:2002np}.  These very specific initial conditions can be ameliorated by including an inflationary regime which will naturally produce a homogeneous and isotropic geometry from a wide variety of initial conditions.  In many brane inflation scenarios, however, only observers on the brane experience inflation and the necessary bulk smoothness remains unexplained.  For the case of two colliding branes, we will show that if $\Lambda_5^{\rm II} > 0$, then the homogeneity and isotropy of both brane and bulk have a natural explanation.  

Prior to the collision, our spacetime consists of two branes and three bulk regions.  Observers living on the left brane see the induced metric from Eq. \eqref{inducedmetric}: 

\begin{equation}
g_{AB}^{\rm ind,L} = n^2(t,vt)\left(\begin{array}{ccc}
-\gamma^2&&v\gamma^2\\&1&\\v\gamma^2&&-v^2\gamma^2\end{array}\right) = (b+(-2d+e+f)vt)^{-2} \left(\begin{array}{ccc}-\gamma^2&&v\gamma^2\\&1&\\
v\gamma^2&&-v^2\gamma^2\end{array}\right) \,.
\end{equation}
A Lorentz transformation diagonalizes this metric, resulting in the 4D metric:
\begin{equation}
g^{\rm ind,L}_{\mu\nu} = (b+(-2d+e+f)v\gamma t')^{-2}\eta_{\mu\nu}\,.
\end{equation}
Finally, a shift in $t'$ eliminates b and gives the conformal time de Sitter metric:
\begin{equation}
g^{\rm ind,L}_{\mu\nu} = \frac{1}{(-2d+e+f)^2v^2\gamma^2{t'}^2}\eta_{\mu\nu} \,,
\end{equation}
and therefore the left brane is generically inflationary with
\begin{equation}
H_-^2 = (-2d+e+f)^2v^2\gamma^2 \,.
\end{equation}
For observers on the right brane, the same arguments give
\begin{equation}
H_+^2 = (-2c-e+f)^2v^2\gamma^2 \,.
\end{equation}
Any initial inhomogeneities on either brane will be inflated away as long as the two brane solution remains valid (i.e. they don't collide) for
$\Delta t \gtrsim \gamma/{\rm Min}\{H_\pm\}$.  We assume that the brane merger can be accomplished in a smooth manner, thus resulting in a homogeneous and isotropic final brane.

Although the branes experience inflation, this is not necessarily the case in the bulk.  For an observer to the left of both branes in region I, the scale factor multiplying the metric takes the form:
\begin{equation}
n_{\rm I}(t,z) = (b+(-c-d+e)z+(c-d+f)vt)^{-1} \,.
\end{equation}
A Lorentz transformation will recast this metric into the standard conformal de Sitter or anti-de Sitter form depending on the relative magnitude of the coefficients multiplying $z$ and $t$.  For $\Lambda_5^{\rm I} < 0$, Eq. \eqref{con1} implies that $\left|-c-d+e\right| > \left|c-d+f\right|v$ and therefore we can make a transformation which eliminates the time dependence.  If a shift in $z$ is included to eliminate $b$, we discover the conformal AdS metric:
\begin{equation}
g^{\rm I}_{AB}(t',z') = \frac{1}{H_{\rm I}^2 {z'}^2}\eta_{AB}\,,
\end{equation}
where $H_{\rm I}^2 = -\Lambda_5^{\rm I}/(6M^3)$.  For positive $\Lambda_5^{\rm I}$, the resulting metric is de Sitter:
\begin{equation}
g^{\rm I}_{AB}(t',z') = \frac{1}{H_{\rm I}^2 {t'}^2}\eta_{AB}\,,
\end{equation}
with $H_{\rm I}^2 = \Lambda_5^{\rm I}/(6M^3)$.
Unsurprisingly, we find that each bulk region will be either dS or AdS depending on the sign of the bulk cosmological constant.

Our desired final state of a flat, stable RS2 universe requires 
$\Lambda_5^{\rm I} = \Lambda_5^{\rm III} < 0$, but $\Lambda_5^{\rm II}$ may take either sign.  If $\Lambda_5^{\rm II} < 0$, all three bulk regions are AdS and initial perturbations may persist through the brane merger and eventually source inhomogeneities on the brane.  A viable cosmology in this case requires a specific set of initial conditions in which those perturbations are absent.  If $\Lambda_5^{\rm II} > 0$, however, the region between the two branes will be inflationary.  Any initial deviations from homogeneity and isotropy will be inflated away in this trapped region.  Once the brane passes through, inflation ends and the spacetime becomes AdS, but it is a very smooth patch of AdS since all initial perturbations have been erased.  When the two branes merge, they will do so in a bulk region that is smooth over a size comparable to their initial separation.  

\begin{figure}[] 
\subfigure[]{
\includegraphics[scale = .7]{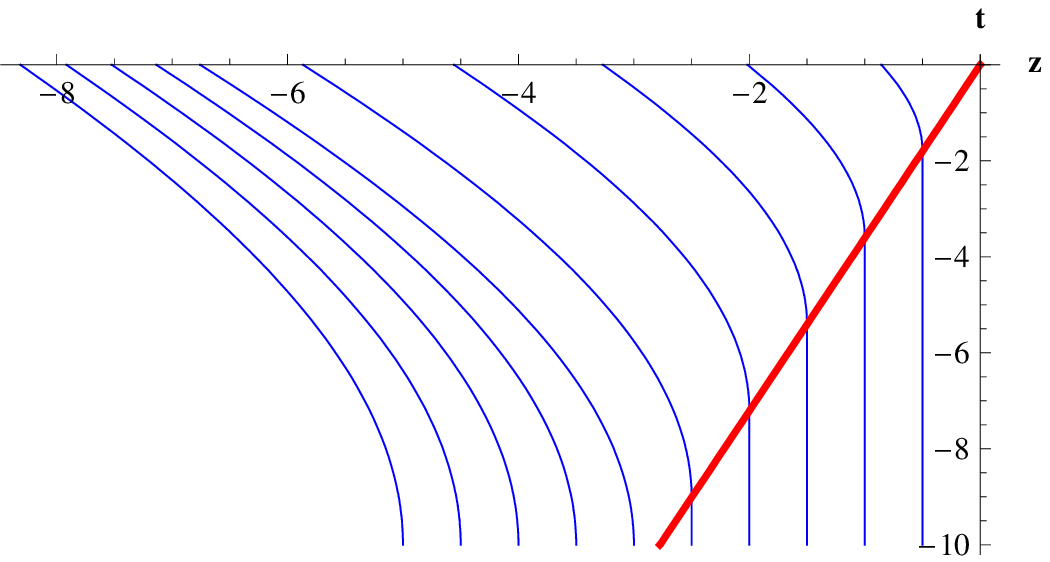}
\label{geo}
} 
\subfigure[]{
\includegraphics[scale = .7]{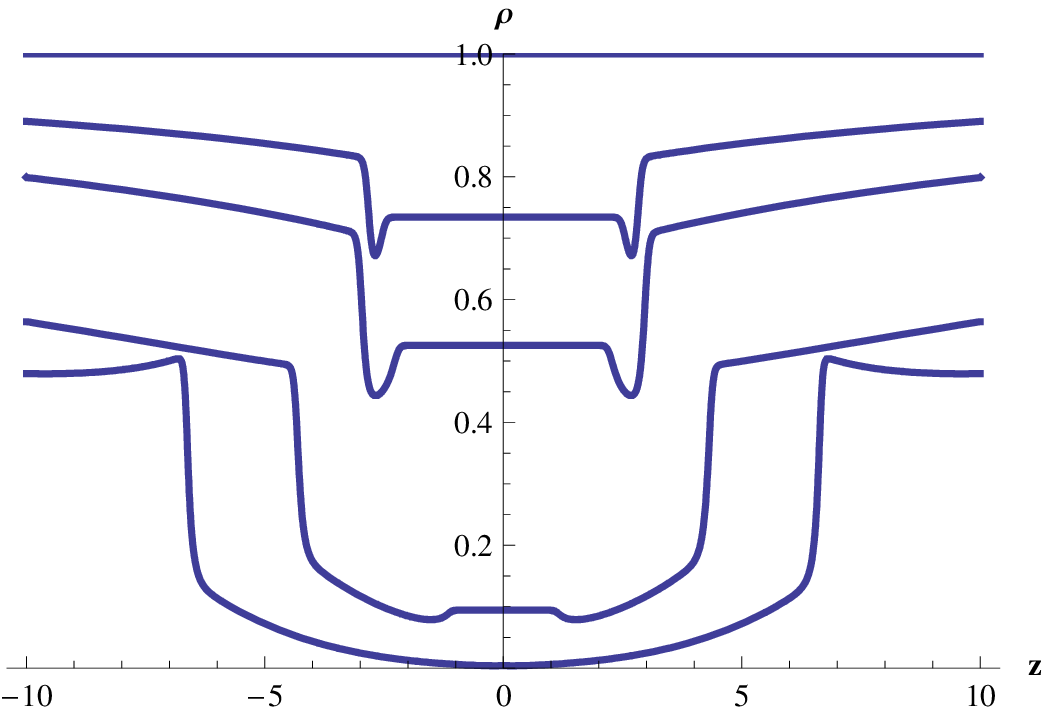}
\label{rho}
}
\caption{Results of numeric evolution for an initially uniform matter distribution in a background spacetime with metric coefficients $b = 1, c = d = \sqrt{13/432}, e = 0, f = -\sqrt{13/27},v= \sqrt{1/13}$., \subref{geo} Geodesics of test particles which are initially stationary at $t=-10$.  The thick diagonal line is the non-geodesic world line of the left brane which collides with its symmetric counterpart at $t = 0$.\subref{rho} Density profile of the matter perturbation at various times.  From top to bottom, the density curves are given at time $t = \{-10,-9,-8,-4,0\}$. }
\label{evol}
\end{figure}

An explicit example of the smoothing properties of this spacetime can be demonstrated by considering the fate of an initial matter perturbation.  For pressureless dust, the stress energy tensor is given by $T^{AB} = \rho U^AU^B$ where $\rho$ is the local energy density and $U^A$ is the five velocity which satisfies the geodesic equation $U^A\nabla_AU^B = 0$.  The evolution of $\rho$ is given by the local conservation of energy:
\begin{equation}
U_B\nabla_AT^{AB} = -\partial_A\rho U^A-\rho\nabla_A U^A = 0 \,.
\end{equation}
For a spacetime with the arbitrarily chosen brane tensions (in units where $M = 1$): $\Lambda_{4-} = \Lambda_{4+} = 1$, $\Lambda_5^{\rm I} = \Lambda_5^{\rm III} = -1/2$, $\Lambda_5^{\rm II} = 1/2$, and the simple initial conditions $\rho(t_0,z) = 1$ and $U^A(t_0,z) = (n^{-1}(t_0,z),0,0,0,0)$, the resultant numeric evolution is displayed in Figure \ref{evol}.  

As expected, by the time of brane collision, the initial matter density has been greatly reduced in the central region.  Just like inflation, this spacetime erases all initial perturbations in a region around the collision point.  Of course, a different choice for the initial matter five velocity will lead to a different evolution for $\rho$, but Figure \ref{geo} reveals an important fact: geodesics in the AdS region are repelled from the brane.  This comes about because, in addition to gravity, the branes feel a force generated by the potential.  As long as $\Lambda_5^{\rm II} > \Lambda_5^{\rm I},\, \Lambda_5^{\rm III}$ this force will be attractive, and the worldlines of particles on the AdS side will diverge from the brane's.  Consequently any geodesics near the point of brane collision most likely originated from the inflationary middle region.  This allows for a wide variety of initial conditions which will evolve into a final state of two smooth branes colliding in a smooth patch of AdS spacetime.  Given the localized nature of gravity in the resultant RS2 universe, it seems reasonable that perturbations far from the brane will have little effect.  A braneworld created in this manner will be well suited to describing our homogeneous and isotropic universe.

\section{Brane merger}\label{branemerger}
In the thin brane limit, we have shown that an inflationary 2 brane solution generally exists for a potential with three or more local minima.  For those potentials which also support a single flat kink solution, we now argue that under certain conditions the two branes will merge to form a flat RS2 universe.  In the weak gravity limit, the merger appears unavoidable.  The dominant force acting on a kink in flat space is given by the change in its potential across the jump \cite{Vachaspati:2006zz}.  In our case, this implies the two branes will feel an attractive force if $\Lambda_5^{\rm II} >
\Lambda_5^{\rm I},\Lambda_5^{\rm III}$.  This force is enhanced for positive brane tensions as Newtonian gravity suggests that an observer at distance $d$ from a brane will experience a constant acceleration toward the brane:
\begin{equation}
\ddot{z} = 4\pi G_5\int_0^\infty\frac{\Lambda_4 r^2 dr}{(r^2+d^2)^{3/2}}
\frac{d}{(r^2+d^2)^{1/2}} = \pi^2 G_5\Lambda_4 \,.
\end{equation}
The 2 branes will be drawn inexorably together into an oscillatory solution.  If any damping occurs (such as scalar radiation) eventually the kinetic energy will be depleted and the branes will be forced to merge.

\begin{figure}[]
\includegraphics[scale = .7]{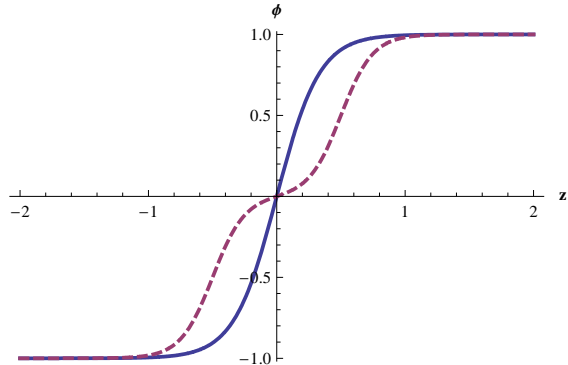} 
\caption{For small velocity and separation, a kink-kink state (dashed) can be considered a perturbation of a single kink state (solid)}\label{kkperturbation}
\end{figure}

More generally, in order for the merger to occur, we require two things: an attractive force between the two branes and a mechanism to shed the brane's kinetic energy.  For kink-kink solutions the first requirement is naturally met by the existence of a stable single kink solution.  As demonstrated in Figure \ref{kkperturbation}, for small velocity and separation, the kink-kink solution can be treated as a perturbation of the single kink solution (this is not true for a kink-antikink solution).  Therefore the force must be attractive since stability guarantees that the end result will be a single brane.  

For large velocities, however, the multi-kink solution can't be considered a perturbation and we need the second requirement.  One possibility is that the final brane tension is roughly equal to the sum of the initial brane tensions and kinetic energies.  As the two near, the kinetic energy can be transformed into potential energy, eventually resulting in the scenario of Figure \ref{kkperturbation}.  However, this requires a fine-tuning in the potential to ensure the appropriate relationship between tensions.  A far more appealing solution is to invoke reheating and dump the excess kinetic energy into particle production on the branes.  Presumably this can be arranged by coupling other matter fields to the scalar field responsible for the branes.  Some initial steps in this direction can be found in Ref. \cite{George:2008vu} which discusses a useful decomposition for fields in a time dependent solitonic background.  An efficient conversion mechanism enables a kink-kink solution moving together at any speed to eventually merge into a single kink solution.

We can explore some of these ideas with an instantaneous brane merger.  Consider the following action:
\begin{align}
S = &-\int \sqrt{-g}\,\left(\Lambda_5^i(t,z)\theta(-t)+\Lambda_5^f(z,t)\theta(t)\right)-\int \sqrt{-g^{\rm ind}}\,\frac{\Lambda_{4-}}{\gamma}\delta(z-vt)\theta(-t) \\
&-\int \sqrt{-g^{\rm ind}}\,\frac{\Lambda_{4+}}{\gamma}\delta(z+vt)\theta(-t) 
-\int \sqrt{-g^{\rm ind}}\,\frac{\Lambda_{4}}{\gamma_2}\delta(z-v_2t)\theta(t) \,, \nn
\end{align}  
where the final brane is allowed a velocity to account for momentum conservation.  We choose the obvious ansatz for $n$:
\begin{equation}
n(t,z) = ((b+c\left|z-vt\right|+d\left|z+vt\right|+ez+fvt)\theta(-t)+
(g+h\left|z-v_2t\right|+jz+kt)\theta(t))^{-1}\,.
\end{equation}
For $t \neq 0$ each part will separately satisfy the constraints found in the first
section.  However, at $t=0$, we need to ensure that the step functions don't add any 
unwanted terms to Einstein's equations.  These terms show up in derivatives of 
$n$ and give the following constraints:
\begin{equation}
b = g \,,\,c+d = h \,,\,e = j\,,\,k = fv \,,\,v_2 = \frac{c-d}{c+d}v \,.
\end{equation}
These restrictions make sense in terms of the tensions.  For the merger to work, the requirements force $\Lambda_5^i = \Lambda_5^f$ on both the left and right hand sides of the final brane - exactly what is needed for thick branes.  In addition, the final brane tension is related to the initial ones by: $\gamma_2\Lambda_4 = \gamma\Lambda_{4-}+\gamma\Lambda_{4+}$.  This is exactly the situation we discussed where the initial kinetic energy is converted into potential energy.  If the effects of reheating are incorporated, we expect this necessary fine tuning to vanish.  More importantly, this simple model demonstrates that a flat brane can arise from the merger of two de Sitter branes. The initial branes were shown to be generically inflationary in Section \ref{inflation}, but a quick calculation reveals that after the merger, the final brane will be flat if $f = e(d-c)/(d+c)$.

The thin brane merger does highlight one other important result: the final brane tension $\Lambda_4$ is proportional to the sum of the individual brane tensions.  For thick branes $\Lambda_4$ measures the size of the jump in $\phi$, therefore this suggests that quick mergers occur for kink-kink initial states but not kink-antikink ones.  There is some agreement for this conclusion in the literature as numeric simulations have been performed for kink-antikink collisions \cite{Takamizu:2006gm,Takamizu:2007ks}.  They show that for relativistic velocities rather than merge, the branes bounce and eventually induce singularities in the spacetime.  These simulations can be extended to study some kink-kink collisions, but not the ones we are interested in, for the method used to set initial conditions guarantees that our desired final state of a flat one-kink solution will not exist.  Regardless, the scenarios outlined in this section suggest that at least for kink-kink initial states a merger into a single brane is quite possible.

\section{Fine tuning}\label{finetuning}
Unlike the standard slow roll inflation in four dimensions, the brane inflation presented here is quite natural, at least at the classical level.  Any appearance of fine tuning in the solutions given in Section \ref{thinbranes} can be directly attributed to our simplifying restrictions to the coordinate system and brane trajectory.  For example, in the weak gravity limit $M \rightarrow \infty$, our solution forces $\Lambda_5^{\rm I} = \Lambda_5^{\rm II} = \Lambda_5^{\rm III}$.  These are the necessary conditions for a constant velocity motion; had we allowed for a more general accelerating solution, this fine tuning would disappear.  We also found that the non-linear nature of Eqs. (\ref{con-1}-\ref{con3}) kept a certain fraction of possible potentials from admitting constant velocity solutions.  We expect these constraints to be loosened in a less restrictive coordinate system.  A concrete example can be given for the special case of two stationary branes.  We consider the action given in Eq.\eqref{action1} with $vt$ replaced by $z_0$.  One set of possible metric solutions is given by the familiar conformally flat metric with
\begin{equation}
n(t,z) = (b+c\left|z-z_0\right|+d\left|z+z_0\right|+ez+ft)^{-1} \,,
\end{equation}
and
\begin{align}
c &= \frac{\Lambda_{4-}}{6M^3} \,, \\
d &= \frac{\Lambda_{4+}}{6M^3} \,, \nn \\
(e-c-d)^2-f^2 &=-\frac{\Lambda_5^{\rm I}}{6M^3} \,, \nn \\ 
(e+c-d)^2-f^2 &= -\frac{\Lambda_5^{\rm II}}{6M^3} \,, \nn  \\
(e+c+d)^2-f^2 &= -\frac{\Lambda_5^{\rm III}}{6M^3} \,. \nn
\end{align}
We see that without $v$, for a given set of $\Lambda$s we now have 5 equations for 4 parameters, and a fine tuning will be required to find a static solution.  For static deSitter branes, however, a more common parameterization is given by
\begin{equation}
g_{AB} = \left(\begin{array}{ccc}-f(z)^2/(H^2t^2)&&\\&f(z)^2/(H^2t^2)&\\&&1\end{array}
\right) \,.
\end{equation}
In a bulk region with negative cosmological constant, the solution for $f(z)$ is given by:
\begin{equation}
f_i(z) = A_i \cosh(m_iz)\pm\sqrt{A_i^2+H^2/m_i^2}\sinh(m_iz) \,,
\end{equation}
where $A_i$ is an arbitary constant and $m_i = \sqrt{-\Lambda_5^{\rm i}/6}$.  Matching conditions at the two brane boundaries give 4 constraint equations.  Since $H$ can be redefined to absorb one of the $A$s, we again appear shorthanded with only 3 free parameters.  However, unlike the previous case, $z_0$ will appear in the constraint equations.  Therefore in this coordinate system, for a given set of $\Lambda$s, a static 2 brane solution will generally exist.  For reasons like this, we expect a complete classification of moving brane solutions to eliminate any need for fine tuning.

Another possibly suspect assumption we have made is that the two branes start out parallel.  At least in the weak gravity limit, though, this assumption arises naturally as the parallel arrangement will be energetically preferred.  In order for the final RS2 universe to be stable, the energy density between the two initial branes must be larger than the energy densities to the sides: $\Lambda_5^{\rm II} > \Lambda_5^{\rm I},\Lambda_5^{\rm III}$.  If we consider the spatial dimensions to have large yet finite extent $L$, the contribution of the middle region to the total energy will be:
\begin{equation}
E^{\rm II} = \Lambda_5^{\rm II}L^3w 
\end{equation}
for parallel branes separated by width $w$, and 
\begin{equation}
E^{\rm II} = \Lambda_5^{\rm II}\frac{L^4}{2}\tan(\theta/2) 
\end{equation}
for branes which intersect at angle $\theta$.  It is clear that for large $L$, parallel configurations will have a much lower energy.  We conclude that no fine tuning is required to append this inflationary regime onto the normal RS2 universe.

\section{Conclusion}\label{conclusion}
The thin brane analysis of the preceding sections strongly support the main claim of this work: that a potential which allows for a single, flat, solitonic thick brane will also admit inflationary multi-brane solutions which can evolve into the single brane configuration.  We have shown this explicitly for the case of two initial branes.  Even with the restrictions we imposed on the coordinate system and brane trajectory, a two brane solution generically exists for an arbitrary potential as Einstein's equations give 5 constraints for 5 unknowns.  This simple counting can be extended to initial states with more than two branes.  Each additional brane adds two constraints, one for the brane tension, and one for the extra bulk cosmological constant.  However, two new parameters will also appear in the metric solution: one, the analogue of $c$ and $d$, and the other, the velocity of the additional brane.  Even for initial states with more than two branes (e.g. a three brane state produced by a double well potential), we find that no tuning of the potential is necessary.

Although multi-kink inflationary states appear inevitable, there is, of course, no guarantee that these configurations will evolve into a single kink final state.  We have argued that a kink-kink merger will take place for small velocities, and that an effective reheating mechanism should induce a kink-kink merger for arbitrary velocities.  A successful kink-antikink merger, on the other hand, may be more difficult to achieve.  For this reason, we believe that the simplest thick brane realization of these ideas will involve the kink-kink initial state.  In addition to constructing actual thick brane solutions, other avenues of potential research include a more complete classification of non-static thin brane solutions, and a calculation of the cosmological perturbations in this model.

\section{Acknowledgments}
We would like to thank M.B. Wise for useful discussions.  This work was supported in part by the U.S. Department of Energy under contract No. DE-FG02-92ER40701.

\section{Appendix}
In the Section \ref{thinbranes}, the constant velocity motion of the branes was inserted by hand.  We will now promote the brane position to a dynamical variable and show that the constant velocity motion is in fact correct.  We start with the Nambu-Goto action for a p-brane:
\begin{equation}
S = -T\int \sqrt{-G}\,d^{p+1}\sigma \,,
\end{equation}
where the integration is over the coordinates on the brane, and the metric on the brane is given by:
\begin{equation}
G_{\mu\nu} = g_{AB}\partial_\mu X^A\partial_\nu X^B \,.
\end{equation}
Here $X^A$ is the worldline of the brane.  We allow for one dynamical degree of freedom in $X^A$ by parameterizing it as:
\begin{equation}
X^A = \left(t,x^1,x^2,x^3,Z(t)\right) \,.
\end{equation}
This will give a kinetic term for the brane.  As before, we will assume that the potential (at least for widely separated branes) only depends on the brane separation.  The extension of our original action to include 2 dynamical branes is given by: 
\begin{align}
S = -&\Lambda_{4-}\int\sqrt{-G^-}\,\delta(z-Z_-(t))d^5x-\Lambda_{4+}\int\sqrt{-G^+}\,
\delta(z-Z_+(t))d^5x \\
-&\int\sqrt{-g}\{\Lambda_5^{\rm I}\theta(-z+Z_-(t))+\Lambda_5^{\rm II}\left(\theta(z-Z_-(t))-\theta(z-Z_+(t))\right)+\Lambda_5^{\rm III}
\theta(z-Z_+(t))\}d^5x \,. \nn
\end{align}
The stress energy tensor for this matter content is found to be:
\begin{align}
T^{AB} = &-g_{AB}\Lambda_5(z,Z_-(t),Z_+(t)) \\
&-g_{AC}g_{BD}\partial_\mu X_-^C\partial_\nu X_-^D
G_-^{\mu\nu}\frac{\sqrt{-G^-}}{\sqrt{-g}}\Lambda_{4-}\delta(z-Z_-(t)) \nn \\
&-g_{AC}g_{BD}\partial_\mu X_+^C\partial_\nu X_+^D
G_+^{\mu\nu}\frac{\sqrt{-G^+}}{\sqrt{-g}}\Lambda_{4+}\delta(z-Z_+(t)) \,.\nn 
\end{align}
If we again choose to parameterize the metric as
\begin{equation}
g_{AB} = n^2(t,z)\eta_{AB}
\end{equation}
then the induced metrics $G_{\mu\nu}^\pm$ take on a simple form:
\begin{equation}
G_{\mu\nu}^\pm = n^2(t,z) {\rm diag}(-1+\dot{Z}_\pm^2,1,1,1) \,,
\end{equation}
and the components of $T_{AB}$ can be calculated.  For the sake of comparison we give $T_{00}$:
\begin{equation}
T_{00} = n^2\Lambda_5+\left|n\right|\frac{\Lambda_{4-}}{\sqrt{1-\dot{Z}_-^2}}
\delta(z-Z_-(t))+\left|n\right|\frac{\Lambda_{4+}}{\sqrt{1-\dot{Z}_+^2}}
\delta(z-Z_+(t))\,.
\end{equation}
We see that if $Z_- = vt$ and $Z_+ = -vt$, then we will have the same $T_{00}$ that we found for the non-dynamical branes.  The same conclusion applies for the other components of $T_{AB}$.  If the brane equation of motion allows for the constant velocity solution, then our non-dynamical results will be valid.  

To find the equation of motion for $Z_-(t)$ we insert our metric ansatz into the action:
\begin{align}
S_{Z_-} = &-\Lambda_{4-}\int\sqrt{n^8(t,z)(1-\dot{Z}_-^2)}\,\delta(z-Z_-)d^5x \\
&-\int n^5(t,z)\{\Lambda_5^{\rm I}\theta(Z_--z)+\Lambda_5^{\rm II}\left(\theta(Z_+-z)-\theta(Z_--z)\right)\}d^5x \,, \nn \\
= &-\Lambda_{4-}\int n^4(t,Z_-)\sqrt{1-\dot{Z}_-^2}\,d^4x \nn \\
&-\Lambda_5^{\rm I}\int\,d^4x\int_{-\infty}^{Z_-} n^5(t,z) dz
-\Lambda_5^{\rm II}\int\,d^4x\int_{Z_-}^{Z_+} n^5(t,z) dz \,. \nn
\end{align}
From this action, we find the equation of motion for $Z_-$ to be:
\begin{equation}
\frac{\Lambda_{4-}n}{(1-\dot{Z}_-^2)^{3/2}}\ddot{Z}_-+\frac{4\Lambda_{4-}}
{\sqrt{1-\dot{Z}_-^2}}\left(\dot{Z}_-n_t+n_z\right) = -\Lambda_5^{\rm I} n^2
+\Lambda_5^{\rm II} n^2 \,,
\end{equation}
where $n$ and its derivatives are evaluated at $z = Z_-$.  Plugging in our proposed solutions for $n$ and $Z_-$, we find what appears to be another constraint:
\begin{equation}
24M^3c(-dv^2+fv^2-d+e) = \Lambda_5^{\rm I}-\Lambda_5^{\rm II} \,.
\end{equation}
However, it turns out that this equation can be derived from the previous constraints by subtracting Eq. \eqref{con2} from Eq. \eqref{con1}.  A similar result applies to the $Z_+$ equation of motion, therefore the branes can be assumed to move at a constant velocity
without any further constraint.

\end{document}